\documentstyle[amssymb,aps]{revtex}

\begin{document}
\author{S. Bruce\thanks{%
E-mail: sbruce@udec.cl, Fax: 56-41-224520.} and L. Roa}
\title{The ground state of a spin-1/2 neutral particle with anomalous magnetic
moment in a Aharonov-Casher configuration}
\address{Department of Physics, University of Concepcion, P.O. Box 4009, Concepcion,\\
Chile.}
\date{March 1999}
\maketitle

\begin{abstract}
We determine the (bound) ground state of a spin $1/2$ chargless particle
with anomalous magnetic moment in certain Aharonov-Casher configurations. We
recast the description of the system in a supersymmetric form. Then the
basic physical requirements for unbroken supersymmetry are established. We
comment on the possibility of {\it neutron trapping} in these systems.

PACS number(s): 03.65Ge, 03.65.Bz, 12.60.J, 11.30.P.
\end{abstract}

Supersymmetric quantum mechanics (SUSYQM) was discussed by Witten as a
laboratory for understanding (super)symmetry breakdown in one-dimensional
field theories \cite{WI1,WI2}. It constitutes a simplified arena where new
ideas are generated, tested, and subsequently generalized. As an example of
a one-dimensional SUSYQM system, let us consider the problem of a chargless
spin $1/2$ particle with anomalous magnetic moment (v.gr. a neutron)
confined to move on the real line under the influence of an axially
symmetric electrostatic field. Aharonov and Casher pointed out the existence
of a quantum mechanical process \cite{AC,K,KA} wherein the behavior of an
uncharged magnetic dipole is affected by the presence of an electric field.
Let us imagine an electrically charged object with axial symmetry centered
around the $z$ axis. The neutrons are completely polarized along, say, the
positive ${\bf y}$ direction. It is straightforward to see that this system
can be recast in a supersymmetric form. In order to study supersymmetry
breaking, we solve the corresponding eigenvalue problem for the ground
state. We conclude that, although in this circumstance there is apparently
no force on the particles, slow neutrons ($s$ states) will tend to move
toward regions where the gradient of the field increases. The next step
considers the standard (two-dimensional) AC configuration. We treat this
problem in the same framework of supersymmetry. Then we perform the
corresponding calculations to determine the physical conditions for unbroken
supersymmetry. Finally we briefly compare our results with those obtained in
the one-dimensional case.

To be specific, let us consider a spin $1/2$ chargless particle with an
anomalous magnetic moment $\kappa _{n}.$ The Dirac equation can be written 
\cite{AC,H,XI} in a covariant form as 
\begin{equation}
\left( \gamma _{\mu }p^{\mu }-\frac{e\kappa _{n}}{2M_{n}}F^{\mu \nu }\sigma
_{\mu \nu }-M_{n}\right) \Psi ({\bf r,}t)=0,  \label{Pa}
\end{equation}
where $F^{\mu \nu }=\partial ^{\mu }A^{\nu }-\partial ^{\nu }A^{\mu }$ is
the electromagnetic field tensor. In 1+1 dimensions we can set 
\begin{equation}
\Psi (z{\bf ,}t)=%
{\psi _{u}(z{\bf ,}t) \choose \psi _{l}(z{\bf ,}t)}%
.  \label{psi}
\end{equation}
Thus from (\ref{Pa}) and (\ref{psi}) we can write the differential equations
for the upper and lower components, $\psi _{u}$ and $\psi _{l},$ as follows 
\begin{eqnarray}
\frac{1}{2M_{n}}\left( p_{z}{\bf +}i\frac{e\kappa _{n}}{2M_{n}}E(z)\right)
\left( p_{z}{\bf -}i\frac{e\kappa _{n}}{2M_{n}}E(z)\right) \psi _{u}(z{\bf ,}%
t) &=&i\frac{\partial \psi _{u}(z{\bf ,}t)}{\partial t},  \label{cou} \\
\frac{1}{2M_{n}}\left( p_{z}{\bf -}i\frac{e\kappa _{n}}{2M_{n}}E(z)\right)
\left( p_{z}{\bf +}i\frac{e\kappa _{n}}{2M_{n}}E(z)\right) \psi _{l}(z{\bf ,}%
t) &=&i\frac{\partial \psi _{l}(z{\bf ,}t)}{\partial t}.  \nonumber
\end{eqnarray}
Assuming that $A^{\mu }$ is time independent, we let the time dependence of $%
\Psi $ be given by 
\begin{equation}
\Psi _{E}(z{\bf ,}t)=\Psi _{E}(z)e^{-iEt}=%
{\psi _{u}(z) \choose \psi _{l}(z)}%
e^{-iEt}.  \label{phi}
\end{equation}
The resulting set of {\it uncoupled} differential equations can be
straightforwardly rewritten in the (non-relativistic) supersymmetric form 
\begin{equation}
{\rm H}_{SS}\Psi _{E}(z{\bf ,}t)=\varepsilon \Psi _{E}(z{\bf ,}t),
\label{su}
\end{equation}
with 
\begin{equation}
{\rm H}_{SS}{\cal =}\left\{ Q,Q^{\dagger }\right\} ,\qquad \left[ {\rm H}%
_{SS},Q\right] =\left[ {\rm H}_{SS},Q^{\dagger }\right] =0,  \label{alg}
\end{equation}
where $\varepsilon \equiv \left( E^{2}-M_{n}^{2}\right) /2M_{n}\geq 0.$ Here 
\begin{equation}
Q\equiv \tau ^{-}\otimes \left( p_{z}{\bf -}i\frac{e\kappa _{n}}{2M_{n}}%
E(z)\right)  \label{sq}
\end{equation}
is the supersymmetric charge with $\tau ^{-}=(1/2)\left( \tau _{1}-i\tau
_{2}\right) ,$ where the $\tau _{1},$ $\tau _{2}$ are Pauli matrices. The
alternative eigenvalue equation (\ref{su}) is indistinguishable from (\ref
{Pa}) (for stationary states) and emerges naturally from it.

From (\ref{su}), (\ref{alg}) and (\ref{sq}) we find 
\begin{equation}
\frac{1}{2M_{n}}\left\{ p_{z}^{2}{\bf -}\tau _{3}\frac{e\kappa _{n}}{2M_{n}}%
\frac{dE{\bf (}z{\bf )}}{dz}+\left( \frac{e\kappa _{n}}{2M_{n}}\right)
^{2}E^{2}(z)\right\} \Psi _{E}(z)=\varepsilon \Psi _{E}(z).  \label{we}
\end{equation}
Thus the {\it superpotential} is proportional to the electric field.

Supersymmetry is unbroken if 
\begin{equation}
Q\Psi _{E=M_{n}}^{(0)}(z)=0,\qquad Q^{\dagger }\Psi _{E=M_{n}}^{(0)}(z)=0,
\label{uc}
\end{equation}
where $\Psi _{E=M_{n}}^{(0)}(z)$ is the normalizable (non-degenerate) ground
state. Without lack of generality we can set 
\begin{equation}
\Psi _{E=M_{n}}^{(0)}(z)\equiv 
{\phi (z) \choose 0}%
,\qquad (s\text{ states}).
\end{equation}
The second equation of (\ref{uc}) is satisfied identically. From the first
one we get 
\begin{equation}
\left( p_{z}{\bf -}i\frac{e\kappa _{n}}{2M_{n}}E(z)\right) \phi (z)=0.
\label{an}
\end{equation}

As a first instance, let us consider the field $E(z)$ created by a uniformly
charged ring of radius $r_{0}$, centered around the origin of the $z$ axis.
From (\ref{an}) we obtain 
\begin{equation}
\phi (z)=C\exp \left( \frac{|\beta |}{\left( z^{2}+r_{0}^{2}\right) ^{1/2}}%
\right) ,  \label{phi1}
\end{equation}
with $\beta \equiv -eQ\kappa _{n}/4M_{n}$, $Q$ the total charge of the
distribution and $C$ a complex constant. Note that when 
\mbox{$\vert$}%
$z|\rightarrow \infty ,$ $\phi (z)\rightarrow C.$ Thus $\phi (z)$ is not
normalizable on the real axis. Therefore $\Psi _{E=M_{n}}^{(0)}$ does not
belong to the Hilbert space, i.e., supersymmetry is broken in this case.

An analogous result is obtained for the case of a disk of radius $r_{0},$
with uniform charge per unit area and total charge $Q$. Here $\phi (z)$
turns out to be 
\begin{equation}
\phi (z)=C^{\prime }\exp \left( -\frac{|\beta |}{\pi r_{0}^{2}}\left(
|z|-\left( z^{2}+r_{0}^{2}\right) ^{1/2}\right) \right) .  \label{phi2}
\end{equation}
Notice that again $\phi (z)$ is not normalizable.

Figure 1 shows the {\it non-normalizable} ground state probability densities 
$\left| \phi \right| ^{2}$ for a neutron in the field of a uniformly charged
ring, and in the field of a uniformly charged disk. Both configurations have
the same total electric charge $Q$.

The third instance regards an infinite plane with charge density per unit
area $\sigma .$ In this example we get 
\begin{equation}
\phi (z)=\sqrt{\frac{|\alpha |}{2}}\exp \left( -\frac{|\alpha |}{2}%
|z|\right) ,  \label{phi3}
\end{equation}
where $\alpha =-e\sigma \kappa _{n}/2M_{n}.$ Here $\phi (z)$ has finite norm
and thus supersymmetry is unbroken. Note that $\phi (z)$ in (\ref{phi2}) and
(\ref{phi3}) is not differentiable at $z=0$ since we have taken an
(idealized) thickless charged surface as a source of the field.

Finally, let us consider an infinitely large uniform charge distribution
with density per unit volume $\rho ,$ where a symmetric infinite plane of
thickness $L$ has been removed. This situation resembles a potential well in
one-dimensional quantum mechanics. In this case we have 
\begin{equation}
\phi (z)=C^{\prime \prime }\exp \left( -\frac{1}{2}\left| \alpha \right|
\left( z^{2}-L\left| z\right| \right) \right) ,\qquad L/2\leq \left|
z\right| ;\qquad C^{\prime \prime }\exp \left( -\frac{1}{2}\left| \alpha
\right| \left( \frac{1}{4}-\frac{1}{2}L\right) \right) ,\qquad \left|
z\right| <L/2,  \label{phi4}
\end{equation}
where now $\alpha =-e\rho \kappa _{n}/4M_{n}.$ Here $\phi (z)$ is also
normalizable and supersymmetry is then unbroken.

In fig. 2, we plot the {\it normalizable} ground-state probability density
of a neutron in the field of a uniformly charged plane. In this picture we
also show the corresponding probability density of an uniformly charged
infinite volume configuration, where a symmetric plane section of thickness $%
L$ has been removed.

Notice that, in the first two examples, confinement mainly comes (although
not completely achieved for the ground state) from the gradient of the
electric field along the $z$ axis. Furthermore, we have found confinement in
one direction {\it assuming} confinement in the other spatial degrees of
freedom.

Next let us examine the standard 1+2 AC configuration \cite{AC,XI}. Here
again we are concerned with the conditions for finding the ground state of a
system with {\it unbroken} supersymmetry. To this end we have to assume {\it %
connectness} in the configuration space in order to be able to define a
normalizable ground state. The problem turns out to have exact supersymmetry
only under the fulfillment of a condition for the magnitude of the charge
distribution which generates the electric field. We also discuss the
possibility of {\it breaking} supersymmetry by examining the requirements
for the existence of lower energy bound states.

To start with, let us consider an infinite cylinder with uniform charge per
unit volume $\rho $ centered along the $z$ axis, so that there exists an
electric field 
\begin{equation}
{\bf E}_{<}{\bf (r)}=\rho {\bf r}/2,{\bf \qquad }0{\bf \leq }r\leq
r_{0};\qquad {\bf E}_{>}{\bf (r)}=\rho r_{0}^{2}{\bf r}/2r^{2}{\bf ,\qquad }%
r_{0}\leq r<\infty ,  \label{E}
\end{equation}
where $r_{0}$ is the radius of the cylinder and for simplicity we have
chosen $\widehat{{\bf r}}{\bf \cdot \widehat{z}=}0.$ Here $\widehat{{\bf r}}$
and $\widehat{{\bf z}}$ are unit vectors in the ${\bf r}$ and ${\bf z}$
directions respectively. The neutrons are completely polarized along the
positive ${\bf z}$ direction. They move on a plane in the the presence of $%
{\bf E.}$ In this circumstance there is no {\it direct} force on the
neutrons but there exists a kind of Aharonov-Bohm effect \cite{AC,K,KA,H}.
Nevertheless, if the singularity on the $z$ axis is removed, as is implied
in (\ref{E}), the neutrons are allowed to penetrate the charged line.
Therefore a new question is to be considered: It regards the problem of the
possible bound states of the neutron in this new AC configuration.

The Aharonov-Casher effective wave equation is obtained by making $A_{0}\neq
0,${\bf \ }${\bf B}={\bf 0,}$ with ${\bf \nabla \cdot E=}\rho $. For
stationary states of energy $E$ we write 
\begin{equation}
\Psi _{E}({\bf r,}t)=%
{\phi ({\bf r}) \choose \chi ({\bf r})}%
e^{-iEt}.  \label{DWF}
\end{equation}
Thus from (\ref{Pa}) and (\ref{DWF}) we get 
\begin{eqnarray}
\frac{1}{2M_{n}}{\bf \sigma }\cdot \left( {\bf p+}\frac{ie\kappa _{n}}{2M_{n}%
}{\bf E(r)}\right) {\bf \sigma }\cdot \left( {\bf p-}\frac{ie\kappa _{n}}{%
2M_{n}}{\bf E(r)}\right) \phi ({\bf r}) &=&\varepsilon \phi ({\bf r}),
\label{ul} \\
\frac{1}{2M_{n}}{\bf \sigma }\cdot \left( {\bf p-}\frac{ie\kappa _{n}}{2M_{n}%
}{\bf E(r)}\right) {\bf \sigma }\cdot \left( {\bf p+}\frac{ie\kappa _{n}}{%
2M_{n}}{\bf E(r)}\right) \chi ({\bf r}) &=&\varepsilon \chi ({\bf r}),
\end{eqnarray}
where ${\bf \sigma =(}\sigma _{1},\sigma _{2})$ and $\varepsilon \equiv
\left( E^{2}-M_{n}^{2}\right) /2M_{n}\geq 0.$ As before, this set of
differential equations can be rewritten in the supersymmetric form 
\begin{equation}
{\rm H}_{SS}{\cal =}\left\{ Q,Q^{\dagger }\right\} ,\qquad \left[ {\rm H}%
_{SS},Q\right] =\left[ {\rm H}_{SS},Q^{\dagger }\right] =0,  \label{HAA}
\end{equation}
with 
\begin{equation}
{\rm H}_{SS}\Psi _{E}({\bf r,}t)=\varepsilon \Psi _{E}({\bf r,}t),
\label{Su}
\end{equation}
where now 
\begin{equation}
Q\equiv \frac{1}{\sqrt{2M_{n}}}\tau ^{-}\otimes {\bf \sigma }\cdot \left[ 
{\bf p-}i\left( e\kappa _{n}/2M_{n}\right) {\bf E(r)}\right]  \label{Sq}
\end{equation}
is the supersymmetric charge and $\tau ^{-}=(1/2)\left( \tau _{1}-i\tau
_{2}\right) ,$ where the $\tau _{1},$ $\tau _{2}$ are Pauli matrices. Thus $%
{\rm H}_{SS}$ is invariant under $N=1$ supersymmetry. Note that the AC
effect has also been discussed in the framework of $N=2$ nonrelativistic
supersymmety \cite{RO}.

From (\ref{ul}) we find \cite{BR} that 
\begin{equation}
\frac{1}{2M_{n}}\left\{ {\bf p}^{2}+\frac{e\kappa _{n}}{2M_{n}}\tau
_{3}\otimes \left( \nabla {\bf \cdot E(r)}+2\sigma _{3}\left( {\bf %
E(r)\times p}\right) _{3}\right) +\left( \frac{e\kappa _{n}}{2M_{n}}\right)
^{2}{\bf E}^{2}({\bf r})\right\} \Psi ({\bf r})=\varepsilon \Psi ({\bf r}).
\label{Sy}
\end{equation}

Supersymmetry is unbroken if 
\begin{equation}
Q\phi ^{(0)}({\bf r})=0,\qquad Q^{\dagger }\phi ^{(0)}({\bf r})=0,
\label{Uc}
\end{equation}
where $\phi ^{(0)}({\bf r})=\phi ^{(0)}(r)$ $\left( r\equiv \mid {\bf r\mid }%
=\sqrt{r_{1}^{2}+r_{2}^{2}\text{ }}\right) $ is the ground state of the
system. In other words, the generators of supersymmetry annihilate the
vacuum state in order to have an exact symmetry. We also have the constraint 
\begin{equation}
\left( {\bf E(r)\times p}\right) _{3}\phi ^{(0)}({\bf r})=\frac{\mid {\bf %
E(r)\mid }}{r}L_{3}\phi ^{(0)}({\bf r})=0\qquad \text{ }(\text{s states}),
\end{equation}
with $L_{3}{\bf =}\left( {\bf r\times p}\right) _{3}$ the $z$ component of
the orbital angular momentum operator. Here we are concerned with states for
which $E^{2}=M_{n}^{2}$, i.e. $\varepsilon =0$.

The second equation of (\ref{Uc}) is satisfied identically since in the
nonrelativistic limit the lower components $\Psi _{E=M_{n}}$ vanish. From
the first one we get 
\begin{equation}
{\bf \sigma }\cdot \left( {\bf p-}i\left( e\kappa _{n}/2M_{n}\right) {\bf %
E(r)}\right) \phi ^{(0)}(r)=0.  \label{Co}
\end{equation}
Without lack of generality we can set 
\begin{equation}
\phi ^{(0)}(r)\equiv 
{\phi (r) \choose 0}%
,\qquad \chi ^{(0)}(r)\equiv 
{0 \choose 0}%
.
\end{equation}
Then from (\ref{Co}) we find the differential equations 
\begin{equation}
\left( \frac{d}{dr}{\bf -}\beta r\right) \phi _{<}(r)=0,\qquad 0\leq r\leq
r_{0};\qquad \left( \frac{d}{dr}{\bf -}\frac{\beta r_{0}^{2}}{r}\right) \phi
_{>}(r)=0,\qquad r_{0}\leq r<\infty ,  \nonumber
\end{equation}
where $\beta \equiv -e\rho \kappa _{n}/4M_{n}.$ Thus 
\begin{equation}
\phi _{<}(r)=Ae^{\frac{1}{2}\beta r^{2}},\qquad 0\leq r\leq r_{0};\qquad
\phi _{>}(r)=Br^{\beta r_{0}^{2}},\qquad r_{0}\leq r<\infty ,
\end{equation}
with $A,B$ complex constants.

Next we demand continuity of the wavefunction and its derivative at $%
r=r_{0}. $ Both conditions give the same information$:A\exp \left[ \left(
1/2\right) \beta r_{0}^{2}\right] =Br_{0}^{\beta r_{0}^{2}}$. Furthermore,
if $\Psi _{E=M_{n}}$ belongs to the Hilbert space, $\phi $ must be
normalizable on the plane $[0,2\pi ]\times \lbrack 0,\infty \lbrack $: 
\begin{equation}
2\pi \int_{0}^{\infty }\mid \phi (r)\mid ^{2}rdr=2\pi \left\{ \mid A\mid
^{2}\int_{0}^{r_{0}}drre^{\beta r^{2}}+\mid B\mid ^{2}\int_{r_{0}}^{\infty
}drr^{2\beta r_{0}^{2}+1}\right\} =1,  \label{N}
\end{equation}
from where we get 
\begin{equation}
\mid A\mid ^{2}=\frac{\mid \beta \mid \left( \mid \beta \mid
r_{0}^{2}-1\right) e^{\frac{1}{2}\mid \beta \mid r_{0}^{2}}}{\pi \left[
2\left( \mid \beta \mid r_{0}^{2}-1\right) \sinh \left( \frac{1}{2}\mid
\beta \mid r_{0}^{2}\right) +\mid \beta \mid r_{0}^{2}e^{-\frac{1}{2}\mid
\beta \mid r_{0}^{2}}\right] }.  \label{No}
\end{equation}
Notice that in (\ref{N}) we must require that 
\begin{equation}
\beta r_{0}^{2}<-1.  \label{Be}
\end{equation}
This inequality constitutes a necessary condition on the possible values of $%
\rho $ and $r_{0}$ (or equivalently on $\lambda \equiv \rho \pi r_{0}^{2}$)
if we want to preserve unbroken supersymmetry. Inserting $c^{2}$ in (\ref{Be}%
), we can estimate the minimum value of $\lambda $ to be able to obtain a
normalizable ground state: $\left| \lambda \right| _{\min }\backsimeq 4\pi
M_{n}c^{2}/\left| e\kappa _{n}\right| \backsimeq 4.\,697\,3\times 10^{-3}$
[C/cm]. As $\lambda $ depends linearly on $r_{0}^{2}$, one can in principle
set up a configuration with the required $\lambda $ \cite{AC}$.$

Figure $3$ shows the neutron density of probability $\mid \phi \mid ^{2}$ as
a function of the dimensionless parameter $r/r_{0}$ for different values of $%
\beta <-1,\ $in natural units. Notice that when $\beta $ approaches $-1$, $%
\mid \phi \mid ^{2}$ becomes flatter, i.e., there exists a larger
probability that the neutron be outside the charged distribution than within
it.

To treat the general eigenvalue problem, we observe that the eigenvalue
problem stated by (\ref{Sy}) has two kinds of solutions: a) non-normalizable
scattering-like states for $\varepsilon >0$ $\left( E^{2}>M_{n}^{2}\right) $%
; b) (normalizable) bound states for $\varepsilon <0$ $\left(
E^{2}<M_{n}^{2}\right) .$

The energy levels are obtained by requiring that the radial solutions and
their derivatives be continuous at $r=r_{0},$ i.e., this is the quantization
condition for the remaining energy levels. This involves non-trivial
numerical calculation and is now being studied. Notice that the existence of
further bound states would break exact supersymmetry, as expressed by (\ref
{Su}) and (\ref{Uc}), since $\left( E^{2}-M_{n}^{2}\right) _{\text{min}}<0$.

From the above we can draw at least two main conclusions: First, in the
one-dimensional systems the electric charge distribution has to be
sufficiently spread out in space in order to preserve unbroken
supersymmetry. If this be the case, $\phi (z)$ is normalizable and thus $%
\Psi _{E=M_{n}}^{(0)}$ constitutes a (unique ground) bound state of the
system. Furthermore, in the standard two-dimensional system, the magnitude
of the electric charge distribution has to be sufficiently large $\left(
\lambda \gtrsim 4\pi M_{n}c^{2}/\left| e\kappa _{n}\right| \right) $ in
order to generate a bound (ground) state. Second, in both the one and
two-dimensional systems, we are not asserting that the neutron {\it directly 
}``feels'' a force due to the electric field generated by the charge
density. Rather, from the second terms on the left hand sides of (\ref{we})
and (\ref{Sy}), we state that the neutron tends to move toward regions where
the {\it gradient} of the electric field increases. The third term in the
same equations corresponds to the appearance of an induced electric dipole
moment on the particle \cite{AC}.

Note that, in the standard AC configuration, the fulfillment of the
condition $E^{2}\leq M_{n}^{2}$ would allow {\it cold neutron trapping} by
an electrostatic field as a physical consequence of a purely quantum
mechanical effect. Confinement is usually achieved by means of diverse
magnetic trap systems \cite{TR}. Cold neutrons are extensively used: in
tests of fundamental quantum theory \cite{AG}, and in applied physics \cite
{EX}.

{\bf Acknowledgments}

This work was supported by Direcci\'{o}n de Investigaci\'{o}n, Universidad
de Concepci\'{o}n, through grants P.I. 96.11.19-1.0 and Fondecyt \#1970995.

One of us (SB) is grateful to the School of Physics, University of
Melbourne, Australia, for its warm hospitality. We are very thankful to
Professors A. G. Klein and G. I. Opat for their valuable criticisms and
helpful suggestions on the experimental and theoretical aspects of this
paper.

FIG. 1. The {\it non-normalizable} ground state probability densities for a
neutron in the field of a uniformly charged ring (blue colour), and of a
uniformly charged disk (green colour); both charge configurations have total
charge $Q$.

\bigskip

FIG. 2. The {\it normalizable} ground state probability density for a
neutron in the field of a uniformly charged plane (blue colour). In green
colour we show the corresponding probability density for a uniformly charged
infinite volume configuration, where a symmetric plane section of thickness $%
L$ has been removed.

\bigskip

FIG. 3. The neutron ground state probability density $\mid \phi \mid ^{2}$
as a function of the dimensionless parameter $r/r_{0}$ for different values
of $\beta <-1$. The units used are $\hbar =c=1$.

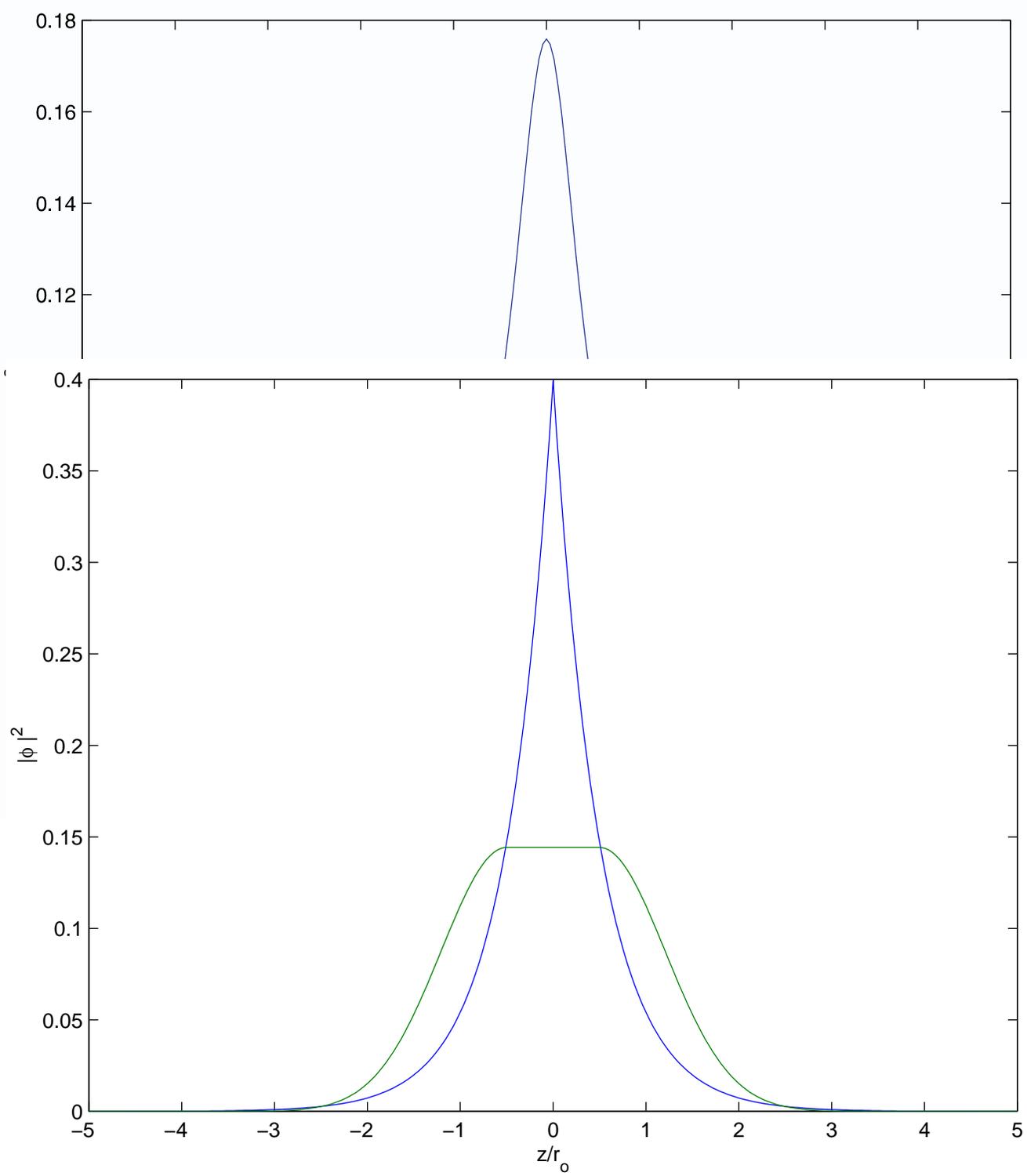
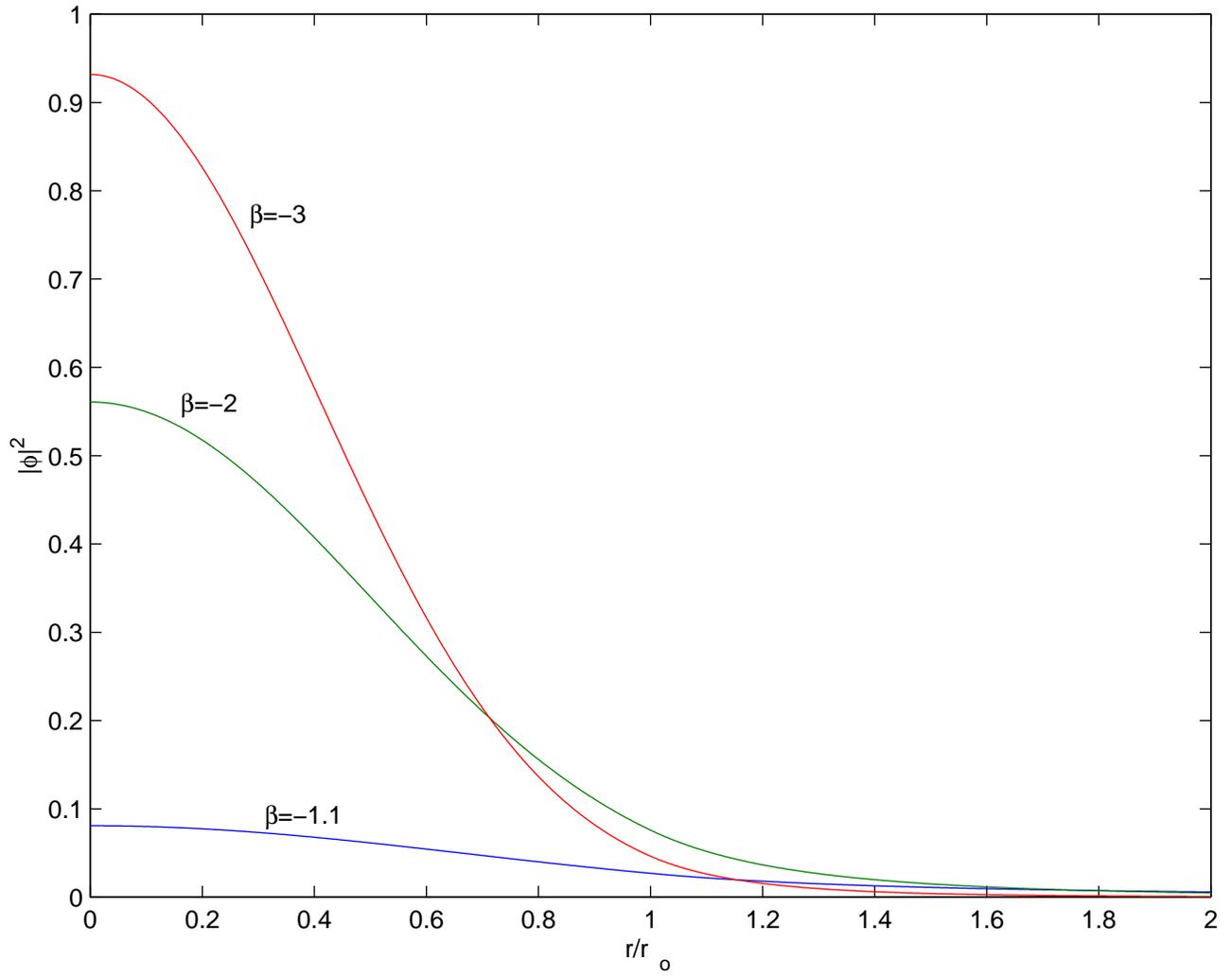
\end{document}